\documentstyle{article}
\title{Bernoulli effect in superconductors and Cooper-pair mass spectroscopy}
\author{M.V. Simkin \\ {\em Department of Physics, Brown University,}\\
{\em Providence, RI 02912-1843}}
\date{ }
\begin{document}
\maketitle
\begin{abstract}
Recently Mishonov (Phys. Rev. B {\bf 50}, 4009 (1994))  suggested to measure 
the Cooper-pair effective
mass using current-induced contact potential difference in superconductors. 
In this Comment it is shown that actual experiments can be substantially
simplified.

74.20.-z, 74.76.-w

\end{abstract}

In his recent articles Mishonov \cite{MISH}, \cite{MISH1} addressed the
intriguing
problem of the measurement of the Cooper-pair mass \cite{TINK}.
It worth noting that he not only justified the possibility of this
measurement like Greiter {\it et al}
\cite{WILC} did before him but he also proposed \cite{MISH} that the Bernoulli
effect in superconductors \cite{LON} should be such a standard tool for this 
measurement
as the Hall effect is for the measurement of charge carrier density in
normal metals. This is a novel point of view as up to now researchers studying
 the 
Bernoulli effect \cite{THEOR},\cite{BOK},\cite{MORIS},
\cite{CHI} where interested only in the effect
itself and did not fully realized its value for the material science.
In the same article  Mishonov \cite{MISH} suggested several modifications
of the experimental methods, used  to measure the Bernoulli effect in
conventional superconductors \cite{BOK},\cite{MORIS},\cite{CHI},
 which would be more appropriate for
layered  High-Tc materials.
The aim of this
Comment  is to propose a substantial simplification of these experiments.

In Ref. \cite{MISH} it was proposed
to make a capacitor of superconducting, insulating and normal metal layers
(see Fig. 1), to place it in an ac magnetic field $B$ , parallel to these 
layers
and measure the Bernoulli voltage between point contacts $(a)$ and $(c)$.
In Ref. \cite{MISH} it was stated that the contact should be made to the 
lateral
surface of the film, where $B=0$. 
It is interesting that in a prototype of this suggested experiment, where 
a superconducting cylinder was placed in a transverse
magnetic field the point contact was indeed made to the place were  $B=0$
(see Fig.1 of Ref.\cite{BOK}).
However in the case of the superconducting film
to make such a contact is not only
technically complicated, particularly for HTSC films, but also
unnecessary. Consider the voltage, as measured by voltmeter,
i.e. the electrochemical potential difference, between $(a)$ and $(b)$, 
$V_{ab}$, and
between $(a)$ and $(c)$, $V_{ac}$. If $V_{ab} \ne V_{ac}$, then $V_{bc} \ne 0$
i.e. the voltmeter will measure nonzero voltage between two parts of 
superconductor in contradiction with experiment \cite{LEW},\cite{BOK},
\cite{MORIS}.

Let us now explain where the voltage $V_{ab}$ comes from. When the magnetic 
field rises from zero to $B$, Meissner currents build up and the Cooper-pairs
at the surface acquire kinetic energy. To make the electrochemical potential 
uniform
the Bernoulli voltage between the surface and the bulk arises. To create this
voltage two surface layers of the film become polarized, while electric field
outside the superconducting film is zero. This means that the electrochemical
potential of the superconducting film changes by the value of kinetic energy
of superconducting electrons on the surface of the film.
At the same time the electrochemical potential of normal metal plate is 
unchanged.
  As a consequence the difference between 
electrochemical potentials of normal metal and superconducting films is equal 
to the kinetic energy of the Cooper-pairs at the surface of the superconducting
film. This kinetic energy can be expressed in terms of the magnetic field at 
the 
surface and Cooper-pairs volume density using London equations. In the case 
when the film is much thicker than the London penetration depth one easily
finds \cite{BOK},\cite{MORIS} that the difference in electrochemical 
potentials is:

\begin{equation}
	V_{ab} = -\frac{B^2}{8\pi n^\ast e^\ast}.
\end{equation}
Here $B$ is the magnetic field at the film surface, $n^\ast$ is the 
Cooper-pairs volume density, $2e^\ast$ is their charge ($e^\ast=\pm e$).
As  was pointed out by Mishonov \cite{MISH}, the sign of the voltage gives
us the sign of the charge carriers (Cooper-pairs), and its magnitude gives us
their volume density, and, as a consequence, their effective mass $m^\ast$(when
one knows $n^\ast$ then $m^\ast$ can be found from the London penetration
depth which depends on $m^\ast/n^\ast$).

Let us now discuss the method of Cooper-pair mass measurement  represented in
Fig. 1 of Ref. \cite{MISH}. From the discussion above it is clear that there
is no need for the ring electrode, while contact (2) can be made to the 
superconducting film surface. This means that the layered 
structure, made by Fiory {\it et al.} \cite{FIO}, can be used as a 
magnetic field induced contact potential difference Cooper-pair
mass spectrometer even without modification suggested in Ref. \cite{MISH}.

I hope that this Comment may encourage experimentalists to measure the
Cooper-pair effective mass using already existing microstructures \cite{FIO},
\cite{X}.

I am grateful to A.Houghton, D.I. Khomskii and J.M. Kosterlitz for useful
discussions.

\newpage

%Figure caption

\begin {figure}
\caption{Magnetic field induced contact potential difference
Cooper-pair mass spectrometer. A capacitor is formed by normal metal (1),
insulating (2), and superconducting (3) layers, grown on the substrate (4).
This structure is placed in a parallel homogeneous ac magnetic field $B$.
This magnetic field induces the alternating difference (on frequency double
to that of magnetic field) in electrochemical potential
between the normal metal and superconducting layers given by Eq.1. 
In the measurement
point contact made to the normal plate (a) and to the surface of the 
superconducting film (b) can be used.
 Point contact (c) to the side surface of the film
is used in discussion only and is not needed in the experiment.}
\end {figure}


\begin{thebibliography}{99}
\bibitem{MISH} T.M. Mishonov, Phys. Rev. B {\bf 50}, 4009 (1994).
\bibitem{MISH1}T.M. Mishonov, Phys. Rev. B {\bf 50}, 4004 (1994).
\bibitem{TINK} M. Tinkham, Introduction to Superconductivity (McGraw-Hill,
New-York,1975)chap.4.1.
\bibitem{WILC} M.Greiter, F. Wilczek, and E. Witten, Mod. Phys. Lett. 
B {\bf 3}, 9003 (1989).
\bibitem{LON}  F.London, Superfluids (John Wiley and Sons Inc., New-York,
1950), Vol.1, p. 58, chap. 8.
\bibitem{THEOR} A.G. Vijfeiken and F.A. Staas, Phys. Lett. {\bf 12}, 175 
(1964),
E. Jakeman and E.R. Pike, Proc. Phys. Soc. {\bf 91}, 422 (1967),
G. Rickaizen, J. Phys. C ser. 2 vol.2, 1174 (1969),
K.M. Hong, Phys.Rev. {\bf B12}, 1766 (1975),
A.N. Omelyanchuk and S.I. Beloborod'ko , Fiz. Nizk. Temp. {\bf 9},1105 (1983)
[Sov. J. Low Temp. Phys. {\bf 9}, 572 (1983)].
\bibitem{LEW} H.W. Lewis, Phys. Rev. {\bf 92}, 1149 (1953);
{\bf 100}, 641 (1955).
\bibitem{BOK} J.Bok and J. Klein, Phys.Rev.Lett. {\bf 20}, 660 (1968).
\bibitem{MORIS} T.D. Morris and J.B. Brown, Physica {\bf 55}, 7609 (1971).
\bibitem{CHI} Yu.N. Chiang and O.G. Shevchenko,  Fiz. Nizk. Temp. {\bf 12},816
 (1986)[Sov. J. Low Temp. Phys. {\bf 12}, 462 (1986)].
\bibitem{FIO} A.T. Fiory {\it et al.}, Phys. Rev. Lett. {\bf65} 3441(1990). 
\bibitem{X} X.X. Xi {\it et al.}, Phys. Rev. Lett. {\bf68} 1240 (1992).
\end{thebibliography}
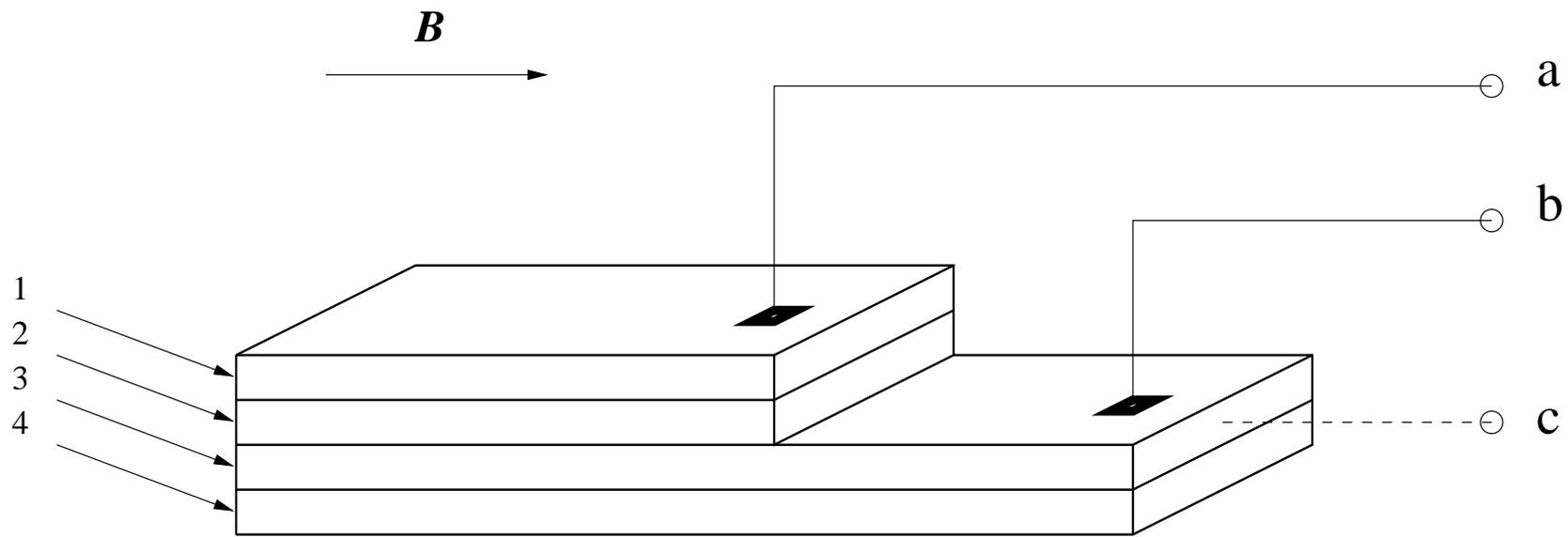
\end{document}